\pdfoutput=1
\documentclass[aip,pof,sd,amsmath,amssymb,reprint,author-year]{revtex4-1}

\usepackage{graphicx}
\usepackage{dcolumn}
\usepackage{bm}

\newcommand{\e}{\ensuremath{\times 10^}}
\newcommand{\partd}[2]{\frac{\partial #1}{\partial #2}}
\newcommand{\diff}[2]{\frac{d #1}{d #2}}

\newcommand{\funcd}[2]{\frac{\delta #1}{\delta #2}}
\newcommand{\ndint}[2]{\mathbf{\mathrm{d}}^{#1} \mathbf{#2}}

\draft 

\begin{document}


\title{Eddy-mixing entropy as a measure of turbulent disorder in barotropic ocean jets} 



\author{Tomos W. David}
\email[]{tomos.david@physics.ox.ac.uk}
\affiliation{University of Oxford, Department of Physics, Clarendon Laboratory, Parks Road, Oxford, OX1 3PU}

\author{Laure Zanna}
\affiliation{University of Oxford, Department of Physics, Clarendon Laboratory, Parks Road, Oxford, OX1 3PU}

\author{David P. Marshall}
\affiliation{University of Oxford, Department of Physics, Clarendon Laboratory, Parks Road, Oxford, OX1 3PU}


\date{\today}

\begin{abstract}
Understanding the statistics of ocean geostrophic turbulence is of utmost importance in understanding its interactions with the global ocean circulation and the climate system as a whole. Here, a study of eddy-mixing entropy in a forced-dissipative barotropic ocean model is presented. Entropy is a concept of fundamental importance in statistical physics and information theory; motivated by equilibrium statistical mechanics theories of ideal geophysical fluids, we consider the effect of forcing and dissipation on eddy-mixing entropy, both analytically and numerically. By diagnosing the time evolution of eddy-mixing entropy it is shown that the entropy provides a descriptive tool for understanding three stages of the turbulence life cycle: growth of instability, formation of large scale structures and steady state fluctuations. Further, by determining the relationship between the time evolution of entropy and the maximum entropy principle, evidence is found for the action of this principle in a forced-dissipative flow. The maximum entropy potential vorticity statistics are calculated for the flow and are compared with numerical simulations. Deficiencies of the maximum entropy statistics are discussed in the context of the mean-field approximation for energy. This study highlights the importance entropy and statistical mechanics in the study of geostrophic turbulence.
\end{abstract}


\maketitle 

\section{Introduction}

Geostrophic turbulence plays an important role in establishing the dynamical balances of the ocean and is an important ingredient in understanding the global ocean circulation \citep[e.g.][]{hallberggnanadesikan2006,marshallj2002,farnetietal2010,farnetidelworth2010,mundayetal2013,marshalletal2017} and its feedback onto the climate system \citep[e.g.][]{lovenduskietal2007,mundayetal2014}. In the Southern Ocean for example, geostrophic turbulence plays a leading order role in setting the transport and stratification for the Antarctic Circumpolar Current \citep{karstenetal2002}. Unfortunately the eddies which make up this turbulence lie beyond the computational reach of modern climate models when run to dynamic and thermodynamic equilibrium. These climate models typically have an ocean resolution of 50-100km while mesoscale eddies have typical length scales of 10-100km, meaning that it is still necessary to parameterize the effect of eddies on the mean flow. Currently the parameterizations typically used in ocean and climate models are variants of that developed by \cite{gentmcwilliams1990} and \cite{gentetal1995} at coarse resolution.

As computational power increases, the resolution of today's climate models is starting to approach the mesoscale therefore permitting a partial representation of the eddy length scales. Although eddy-permitting resolution is advantageous in terms of beginning to resolve eddy dynamics, it may be the case that is becomes no longer appropriate to use a deterministic parametrization such Gent-McWilliams. As resolutions approach the mesoscale the average effect of sub-grid eddies may not be representative as there will be an insufficient number of eddies below the sub-grid scale for averaging to be meaningful. An alternative approach is to model eddies with a stochastic parameterization. Recent studies have explored this possibility \citep{berloff2005,manazanna2014,groomsetal2015,zannaetal2017} for ocean models. With this growing interest in the stochastic nature of mesoscale eddies, it is timely to study the statistics of potential vorticity, and the underlying organizing principles influencing these statistics, in simplified ocean models. One approach is to borrow concepts from statistical physics such as entropy and equilibrium statistical mechanics. This approach has a long history which goes back to the birth of turbulence theory.   

The first example of the application of statistical mechanics to two-dimensional turbulence comes from \cite{onsager1949}, where a model of singular point vortices was proposed. The statistical mechanics of point vortices has received much study since then. \cite{kraichnanmontgomery1980} led the way towards continuous vorticity fields through utilizing the invariance of energy and enstrophy in variational problems. Geostrophic flows over topography were tackled using this methodology: in \cite{salmonetal1976} via the maximization of an entropy; and independently in \cite{brethertonhaidvogel1976} via a phenomenological minimum enstrophy principle. 

In the early 1990s, work was published which led to a theory of equilibrium statistical mechanics of two-dimensional and geophysical flows \citep{miller1990,robert1990,robert1991,robertsommeria1991,milleretal1992}. We refer to this theory as the `Miller-Robert-Sommeria statistical mechanics'. Reviews of this field are given by \cite{chavanis2009} and \cite{bouchetvenaille2012}. The power of the Miller-Robert-Sommeria theory is that all the work of \cite{onsager1949}, \cite{kraichnanmontgomery1980}, \cite{salmonetal1976} and \cite{brethertonhaidvogel1976} are contained within this framework as particular limits or simplifications. This ideal (no forcing, no dissipation) theory has been used to suggest a statistical explanation for the formation of ocean rings and jets in \cite{venaillebouchet2011}. A further example of equilibrium statistical mechanics ideas being applied to the problem of ocean circulation and its associated density profiles is presented by \cite{salmon2012}. 

In this study we consider the concept of entropy in the context of dissipative and forced-dissipative ocean flow. The powerful theory of equilibrium statistical mechanics is limited in usefulness when tackling non-ideal fluids; incorporating the effect of forcing and dissipation on entropy is essential in addressing the extent to which entropic ideas can be applied to the problem of mesoscale ocean turbulence. The aims of this study are as follows.
\begin{itemize}
\item To determine the impact of forcing and dissipation on the evolution of entropy in a turbulent barotropic jet, analytically and numerically.
\item To test the maximum entropy principle\footnote{Not to be confused with the similarly named maximum entropy production principle.} and to understand the utility of this principle in the context of an idealized forced-dissipative turbulent jet.
\item To use the maximum entropy principle as a means to formulate a relationship between dynamically balanced quantities and the statistics of the flow. 
\end{itemize}

In Section \ref{modelexp}, we outline the barotropic model used in this study and describe the numerical experiments. In Section \ref{entdef}, we introduce the eddy-mixing entropy and define the key concepts in this study. In Section \ref{analent}, we derive analytical expressions for the influence of constant forcing and linear drag on the  entropy as well as consider the case of a Gaussian stochastic forcing. In Section \ref{freedecay}, we diagnose the entropy for a freely-decaying turbulent jet and use this to test the predictions of Section \ref{analent}; in Section \ref{force-diss}, we repeat the diagnosis for a forced-dissipative turbulent jet and consider entropy as balanced dynamical quantity. In Section \ref{maxent}, we consider the maximum entropy principle and relate it to the ideas developed in the subsequent parts of this study, evidence is given in support of the maximum entropy principle and we discuss this in relation to forced-dissipative flow. In addition the mathematical problem of obtaining the statistics of the flow from the assumption of maximum entropy is presented and the maximum entropy statistics are calculated and compared with simulations. In Section \ref{discmeanfield}, we discuss the mean-field approximation for energy and its relation to the maximum entropy statistics. In Section \ref{conc}, the study is concluded with some closing remarks. 

\section{Model and experiments}
\label{modelexp}
\subsection{Model}

We solve the barotropic vorticity equation on a $\beta$ plane within a singly-periodic domain, recently used in \cite{davidetal2017a}. The simplicity of this model allows us to perform many high resolution simulations and the channel configuration provides an analogue to the Southern Ocean. The equation of motion is given by
\begin{equation}\label{eom}
\partd{q}{t} = - \lbrace \psi, q \rbrace - r \nabla^2 \psi - \nu_h \nabla^6 \psi - \partial_y \tau(y),
\end{equation}
where the potential vorticity is given by 
\begin{equation}
q = \nabla^2 \psi + \beta y;
\end{equation}
$\nabla^2 \psi$ is the relative vorticity and $\beta y$ is the planetary vorticity making $q$ equivalent to the absolute vorticity in this barotropic model; $\tau$ is the zonal wind stress which is defined to be a function of meridional distance only; $r$ and $\nu_h$ are the linear drag coefficient and the hyper-viscosity respectively; the braces denote the horizontal Jacobian operator. The biharmonic diffusion is used for numerical stability, preferentially dissipating the grid-scale noise compared to Laplacian diffusion. The hyper-viscosity is chosen to be as small as possible while allowing us to treat the linear drag as the dominant dissipative term in this study. Linear drag is an attractive choice of dissipative term due to its analytical tractability as well as its being analogous to oceanic bottom drag.

The periodicity in the zonal direction is employed to solve the model using a pseudo-spectral method and is modified from a pre-existing code \citep{esler2008,eslerhaynes1999}. The model domain is shown in Figure
\begin{figure}[ht!]
\centerline{\includegraphics[width=7cm]{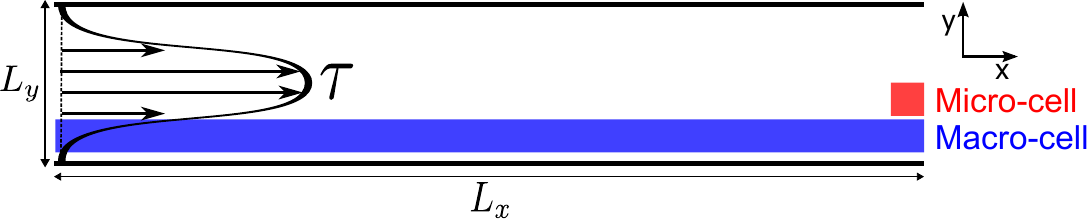}}
\caption{Geometry of domain. wind stress, $\tau$, in this study points from left to right. $L_x$ and $L_y$ are the longitudinal and latitudinal extents of the domain respectively. Macro-cells (blue) and micro-cells (red) in this study. Exploiting zonal symmetry zonal bands are used as macro-cells whilst micro-cells are the grid-points within.}
\label{macromicrodomain}
\end{figure}
\ref{macromicrodomain}. The boundary conditions are free-slip
\begin{equation}
\nabla^{2n}\psi|_{N,S} = 0,
\end{equation}
where $n=1,3$; no normal flow
\begin{equation}
\partial_x \psi|_{N,S} = 0;
\end{equation} 
and global momentum conserving
\begin{equation} \label{momconscond}
\psi|_{N,S} = \pm \frac{\Gamma(t)}{2}.
\end{equation}
We find $\Gamma$ by solving the prognostic integral momentum balance,
\begin{equation}
\diff{\Gamma}{t} = -\iint  \ndint{2}{x} \left[ r \partd{\psi}{y} + \tau \right].
\end{equation}
This is the same condition used by \cite{treguier1989} and is the barotropic (and rigid lid) limit of the general integral momentum balance derived in \cite{mcwilliams1977}. By applying this boundary condition we are able to impose a fixed wind stress forcing rather than relaxing to a background shear as is often done for models of this type \citep[e.g.][]{phillips1954}. 

It is important to note that the ideal dynamics of this flow conserves the following quantities in addition to momentum. 

In ideal flow the energy
\begin{equation}
E = \frac{1}{2} \iint \ndint{2}{x} \, (\nabla \psi) \cdot (\nabla \psi)
\end{equation}
is conserved. This can be rewritten as 
\begin{equation}
E = \frac{\Gamma}{4} (u_S - u_N) - \frac{1}{2} \iint \ndint{2}{x} \, \psi (q - \beta y),
\end{equation}
exploiting the relationship between $q$ and $\psi$ as well as the boundary condition described above, where $u_N$ and $u_S$ are the velocities along the north and south boundaries respectively. When the flow exhibits symmetry breaking this term will be non-zero, this is not the case in the flow realizations presented in this study. 

In addition, ideal flow conserves the integral of any function of potential vorticity
\begin{equation}
C = \iint \ndint{2}{x} \, c(q),
\end{equation}
where $c$ is an arbitrary function. These quantities are called Casimirs \citep[e.g.][]{salmon}. We are primarily interested in the polynomial Casimirs which we will denote as 
\begin{equation}
C_n = \iint \ndint{2}{x} \, q^n.
\end{equation}
Two Casimirs of particular physical importance are the circulation, $n=1$, and the enstrophy, $n=2$. Alternatively, all Casimirs can be conserved simultaneously by conserving the global potential vorticity distribution, $\Pi$, where
\begin{equation}
\Pi(q) = \diff{A(q)}{q}
\end{equation}
where $A(q)$ is the area of the domain occupied by points with a value of potential vorticity less than $q$, the global cumulative potential vorticity distribution function.

\subsection{Experiments}

Two numerical experiments are performed using the above model. 

The first experiment is a freely-decaying unstable jet in which the initial jet has a velocity profile
\begin{equation}
u(y) = u_0 \mathrm{sech}^2(y),
\end{equation} 
with $u_0=10$. The unstable jet evolves freely under the action hyper-viscosity and varying strengths of linear drag. 

The second experiment is a forced dissipative turbulent jet spun up from rest with varied strength of the wind stress. The wind stress profile is kept the same for all simulations as,
\begin{equation}
\tau = \tau_0 \mathrm{sech}^2\left(\frac{y}{\delta}\right),
\end{equation} 
but the magnitude of the jet is varied by changing the value of $\tau_0$. $\delta$ is the width parameter and is fixed for all simulations. Table 
\begin{table}
\begin{center}
\begin{tabular}{cccc}
\hline
Simulation  & Wind stress & Linear drag   & Timestep, $\mathrm{d}t$ \\ 
ID & strength, $\tau_0$ &  coefficient, $r$ & \\ \hline
\multicolumn{4}{l}{\emph{Freely-decaying experiments}} \\
$\mathrm{D_1}$ & 0.000 & 0.0008 & 0.0005\\
$\mathrm{D_2}$ & 0.000 & 0.0009 & 0.0005\\
$\mathrm{D_3}$ & 0.000 & 0.0010 & 0.0010\\
$\mathrm{D_4}$ & 0.000 & 0.0020 & 0.0010\\
$\mathrm{D_5}$ & 0.000 & 0.0030 & 0.0010\\
$\mathrm{D_6}$ & 0.000 & 0.0040 & 0.0010\\
$\mathrm{D_7}$ & 0.000 & 0.0050 & 0.0010\\
\multicolumn{4}{l}{\emph{Forced-dissipative experiments}} \\
$\mathrm{FD_1}$ & 0.005 & 0.0050 & 0.0010\\
$\mathrm{FD_2}$ & 0.010 & 0.0050 & 0.0010\\
$\mathrm{FD_3}$ & 0.020 & 0.0050 & 0.0010\\
$\mathrm{FD_4}$ & 0.040 & 0.0050 & 0.0010\\
$\mathrm{FD_5}$ & 0.080 & 0.0050 & 0.0010\\
$\mathrm{FD_6}$ & 0.160 & 0.0050 & 0.0010\\
$\mathrm{FD_7}$ & 0.320 & 0.0050 & 0.0010\\

\end{tabular}
\caption{List of experiments and non-dimensional parameters.} \label{runs}
\end{center}
\end{table}
\ref{runs} gives the values used in the different simulations. The parameters which are held constant for both experiments are given in Table
\begin{table}
\begin{center}
\begin{tabular}{ll}
\hline
Parameter  & Value \\ \hline
Meridional extent, $L_y$ & $5\pi/2$ \\
Zonal extent, $L_x$ & $20 \pi$ \\
Number of zonal grid-points, $n_x$ & $1024$ \\
Number of meridional grid-points, $n_y$ & $128$ \\
Time-step, $\mathrm{d}t$ & $1\e{-3}$\\
Output frequency & 1 \\
Total time of output & $1\e{4}$ \\
Hyper-viscosity, $\nu_h$ & $2\e{-6}$ \\
Wind stress width parameter, $\delta$ & $0.4$ \\
Beta parameter, $\beta$ & $0.2$ 
\end{tabular}
\caption{Fixed parameters of model simulations.} \label{param}
\end{center}
\end{table}
\ref{param}.

\section{Eddy-mixing entropy}
\label{entdef}

The eddy-mixing entropy is not the same as the thermodynamic entropy associated with molecular motions. The eddy-mixing entropy is a measure of the disorder of the large scale turbulent flow, and depends on the choice of coarse-graining which distinguishes between the large scales and the small scales of the flow. To formally define the eddy-mixing entropy we follow the approach presented in \cite{jungetal2006}. To proceed we will define two sub-systems of the full flow:
\begin{itemize}
\item A \emph{micro-cell} which is the smallest scale over which the details of the flow are important. The micro-cell is equivalent to the grid-cell for a high resolution numerical simulation. We think of each micro-cell as being characterized by only one value of the potential vorticity.
\item A \emph{macro-cell} which is comprised of a number of micro-cells and is related to a choice of some coarse graining scale. The macro-cells should be chosen to exploit some dynamical symmetry of the system. In our case we, for the most part, choose zonal bands exploiting the zonal symmetry of the system apart from in Section \ref{recentevo} where we choose contours of instantaneous streamfunction.
\end{itemize}
The macro- and micro-cells used in this study are schematically illustrated in Figure \ref{macromicrodomain}. 

Having defined the marco-cells as such, it is possible to define an eddy-mixing entropy by counting the number of ways to arrange the micro-cells of value of potential vorticity into the macro-cells. This gives the expression
\begin{equation}\label{gibbsmixingentropy}
S = \ln W = \ln \prod_I \frac{M^{(I)}!}{\prod_r M^{(I)}_r !}
\end{equation}
for the eddy-mixing entropy, where $M^{(I)}$ is the number of micro-cells in the $I^{\mathrm{th}}$ macro-cell and $M^{(I)}_r$ is the number of micro-cells with the $r^{\mathrm{th}}$ value of potential vorticity in the $I^{\mathrm{th}}$ macro-cell. This counting method is adapted from \cite{lyndenbell1967} where is was used to study the statistical mechanics of stellar systems. 

For large numbers of micro- and macro-cells we can take the continuous limit to get 
\begin{equation}\label{eddymixingentropy}
S[\rho]=- \int \ndint{2}{x} \mathrm{d}\tilde{q} \, \rho(\tilde{q} |\mathbf{x}) \ln( \rho(\tilde{q} |\mathbf{x})),
\end{equation}
in terms of the probability distribution function, $\rho$. In words, $\rho$ \emph{is the probability of measuring a value, $\tilde{q}$, of the potential vorticity at the point $\mathbf{x}$ in the domain.} It is important to note that in this expression $\mathbf{x}$ has taken the place of $I$ in labelling the macro-cell. The coordinate $\mathbf{x}$ should be interpreted as a coarse-grained or smoothed coordinate. The $\tilde{q}$ is not the potential vorticity field, $q(\mathbf{x})$, but a random variable representing the result of a measurement of potential vorticity. The eddy-mixing entropy is the sum over the continuous information entropies associated with the distribution of potential vorticity in each macro-cell. The method of numerically determining these entropies are given in an appendix and is used throughout this study.

\section{Analytical model for evolution for entropy}
\label{analent}

The fundamental quantity that we are interested in is the eddy mixing entropy given by \eqref{eddymixingentropy}. In this section we derive a tendency equation for this entropy. We are not able to derive a full theory as the effects of the non-linear or non-local terms in the vorticity equation, \eqref{eom}, do not seem to be analytically tractable.  Nevertheless it is possible to derive analytical expressions for the entropy evolution due to the remaining linear and local terms in the vorticity equation: the wind stress curl and the linear drag. To allow for wider applicability to future studies, we also include a Gaussian stochastic noise, although we do not consider the stochastic forcing case in the numerical experiments described in this study. 

Ignoring the non-local (hyper-viscous) and non-linear (advection) terms we have the following equation for the evolution of potential vorticity:
\begin{equation}\label{lineom}
\partd{q}{t} = -r (q - \beta y) - g(y) + \eta
\end{equation}
where $g(y)=-\partial_y \tau(y)$ is the constant (in time) part of the forcing and $\eta$ is a Gaussian noise with zero mean and is uncorrelated in time. The stochastic part of the forcing might be considered as an analogue to the high frequency part of the atmosphere's interaction with the ocean as well as high frequency ocean processes not represented in this model. 

Equation \eqref{lineom}, ignoring the non-local and non-linear terms, represents a Ornstein--Uhlenbeck process \citep[e.g.][]{gardiner} and the corresponding probability distribution satisfies the Fokker-Planck equation,
\begin{equation}\label{fokkerplanck}
\partd{\rho}{t} = \partd{}{\tilde{q}} [r(\tilde{q} - \beta y)\rho] - g(y)\partd{\rho}{\tilde{q}} + \frac{\langle \eta^2 \rangle}{2} \partd{^2 \rho}{\tilde{q}^2},
\end{equation}
where $\langle \eta^2 \rangle$ is the variance of the noise. The fist two terms on the right hand side of equation \eqref{fokkerplanck} were also derived in \cite{kazantsevetal1998} but here we consider the effect of these terms on the entropy. We can write the entropy tendency as 
\begin{equation}\label{sdot}
\dot{S} = -\int \ndint{2}{x} \mathrm{d}\tilde{q} \, \dot{\rho} \ln \rho, 
\end{equation}
where the dot represents differentiation with respect to time. By substituting equation \eqref{fokkerplanck} into \eqref{sdot} we can derive the influence of these terms on the entropy, yielding 
\begin{equation}\label{entevofull}
\dot{S} = P -r + \frac{\langle \eta^2 \rangle}{2} \int \ndint{2}{x} \mathrm{d}\tilde{q} \, \partd{^2 \rho}{\tilde{q}^2} \ln \rho,
\end{equation}
where we have now included the effect of advection and hyper-viscosity as the residual, $P$; as the hyper-viscosity is small in our numerical calculations, we will take the liberty of referring to $P$ as the \emph{advective production of entropy}. 

Notably, \eqref{entevofull} does not have an explicit dependence on the zonally symmetric forcing as a constant wind stress only shifts the distribution in each zonal band and the entropy is invariant to these shifts. The linear drag leads to a remarkably simple term which is minus the drag coefficient, that is a perpetual and constant sink of entropy. Thus, any increase in entropy must arise through the effect of turbulence or stochastic forcing. In the absence of turbulence, the stochastic noise term will solely act to spread out the probability distribution and will consequently act as a source of entropy. In this case we would expect there to be a stationary solution for the entropy evolution in the case of no eddies where the source of entropy due to stochastic noise balances the sink of entropy due to linear drag. This can be tested by numerically solving the linear equation of motion, \eqref{lineom}, and comparing with the steady state solution of the Fokker-Planck equation, \eqref{fokkerplanck}, and its corresponding entropy. In what follows we will not consider the case of Gaussian stochastic forcing but will concentrate of simulations with constant zonally symmetric forcing. We include the above as an important caveat to the numerical results presented in this study; in a realistic ocean simulation high frequency noise will introduce a source of entropy from its fast time-scale variability. 

\section{Numerical experiments}
\subsection{Eddy-mixing entropy in freely-decaying turbulence}
\label{freedecay}
\begin{figure*}[ht!]
\centerline{\includegraphics[width=\textwidth]{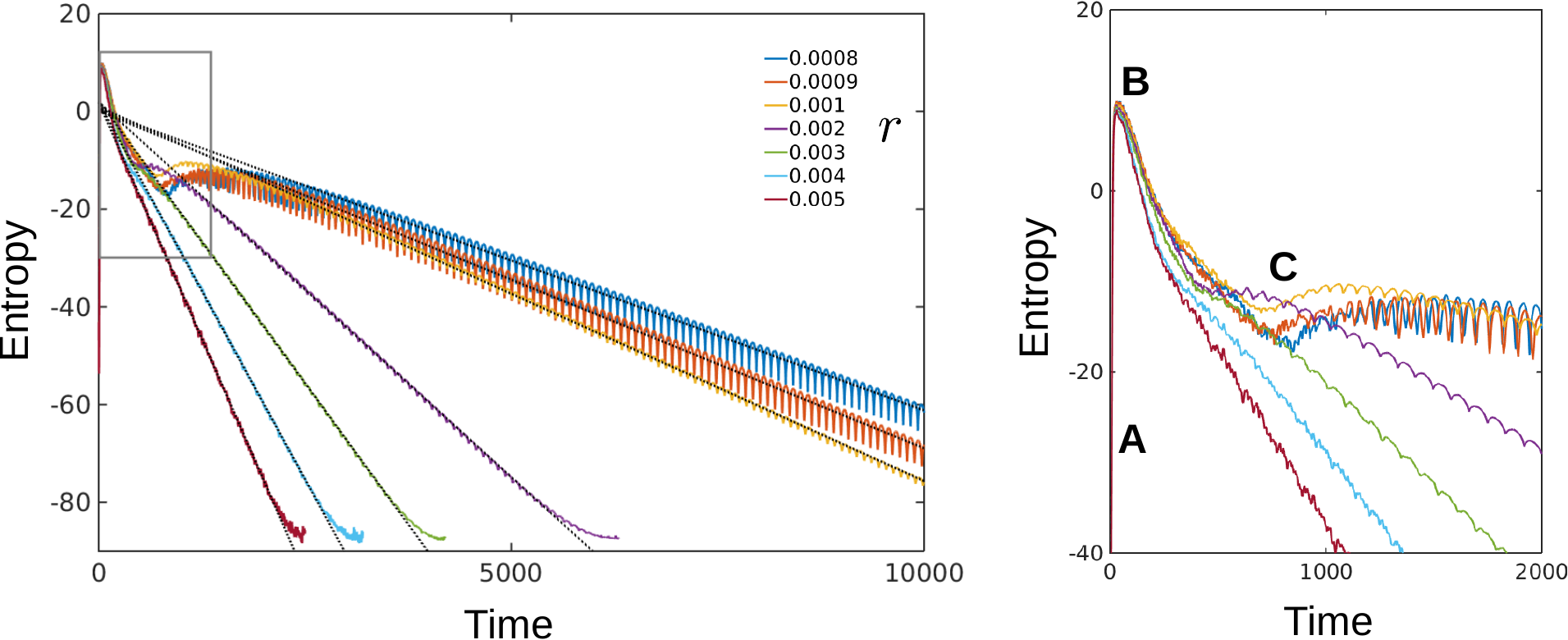}}
\caption{Entropy as a function of time for different values of linear drag coefficient, simulations $\mathrm{D_1,...,D_7}$. A - We find that exponential growth of barotropic instabilities lead to a very fast growth in the entropy. B - The entropy reaches a maximum value which is insensitive to the value of the linear drag coefficient. This is followed by a decrease in entropy at a rate greater than the contribution of the linear drag. This implies that this decrease is due to the eddies themselves. C - At long time the entropy decay is linear and is explained by the linear drag. For the lowest values of linear drag coefficient a persistent Rossby wave emerges causing a secondary increase in the entropy before the long-time behaviour is seen.
}
\label{decay}
\end{figure*}
For our numerical simulations (no stochastic term), the entropy tendency is simply given as  
\begin{equation}\label{entevo}
\dot{S} = P - r,
\end{equation}
with no explicit contribution from the deterministic forcing term. However, the forcing will contribute to determining the behaviour of the advective production of entropy, $P$. To illuminate the effect of forcing we first consider the evolution of entropy in the absence of forcing: the freely decaying unstable flow.

We begin by examining the evolution of entropy for short times as the instabilities grow then decay shown in Figure \ref{decay}. In region A of Figure \ref{decay}, the entropy increases very quickly concomitant with the exponential growth of eddy energy through shear instability in the jet. There is little spread in rate of the entropy growth in simulations $\mathrm{D_1,...,D_7}$ with changed drag parameter. This growth is arrested for all experiments at a maximum value in region B, where the maximum is also not greatly changed with the differing linear drag coefficient. The entropy then decreases, in regions B to C, towards its asymptotic behaviour. This decrease is \emph{greater} than can be explained by the sink of entropy due to linear drag meaning that in this region, B to C, the advective production of entropy must become negative and acts as a sink of entropy. 

There is an interesting difference between simulations $\mathrm{D_1,...,D_3}$ and the other simulations. These low drag simulations see a second increase in entropy (Figure \ref{decay}, near $\mathrm{Time} = 1000$) toward the long-time behaviour as well as an oscillation about the long time behaviour. These effects can be illuminated by examining the flow, at low linear drag coefficient a persistent Rossby wave forms causing an increase in the disorder as compared to laminar flow as well as the observed oscillations. 

In freely-decaying turbulence the eddies will ultimately die away through the action of linear drag and hyper-viscosity causing the advective production of entropy, $P$, to tend to zero at long times. In this case equation \eqref{entevo} tends to the asymptotic solution for the entropy evolution 
\begin{equation}\label{entdecaypredict}
S(t) \approx -r t + K \quad \text{for long times},
\end{equation}
where $K$ is a constant of integration. We can test this hypothesis in a simulation of a freely decaying unstable jet as for long times we would expect the activity of the eddies to asymptotically decay to zero. We initialize the model with an unstable zonal jet and allow the flow to evolve under the action of linear drag and hyper-viscosity only, the latter being small.  

The long time behaviour of the entropy is also shown in Figure \ref{decay}. We see a striking agreement between the long time behaviour predicted by \eqref{entdecaypredict} and the slopes diagnosed from the simulations. Figure
\begin{figure}
\centerline{\includegraphics[width=7cm]{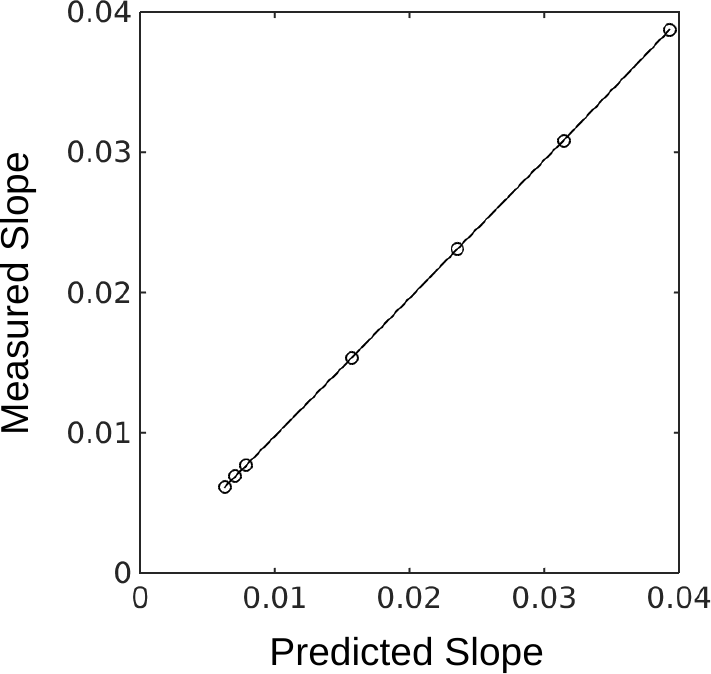}}
\caption{Measured slope of long time linear decay of the entropy, diagnosed from simulations $\mathrm{D_1,...,D_7}$, plotted against the predicted slope. The agreement is near perfect.}
\label{predvsmeas}
\end{figure}
\ref{predvsmeas} shows the agreement between the predicted and the measured long time slope of the linear entropy decrease which is found to be near exact.

\subsection{Eddy-mixing entropy in forced-dissipative turbulence}
\label{force-diss}
We now turn our attention to the entropy evolution in the forced-dissipative experiments $\mathrm{FD_1,...,FD_7}$. We start by considering the entropy for short times, comparing it to snapshots of the potential vorticity. As an example we consider experiment $\mathrm{FD_3}$. Figure
\begin{figure}[t!]
\centerline{\includegraphics[width=7cm]{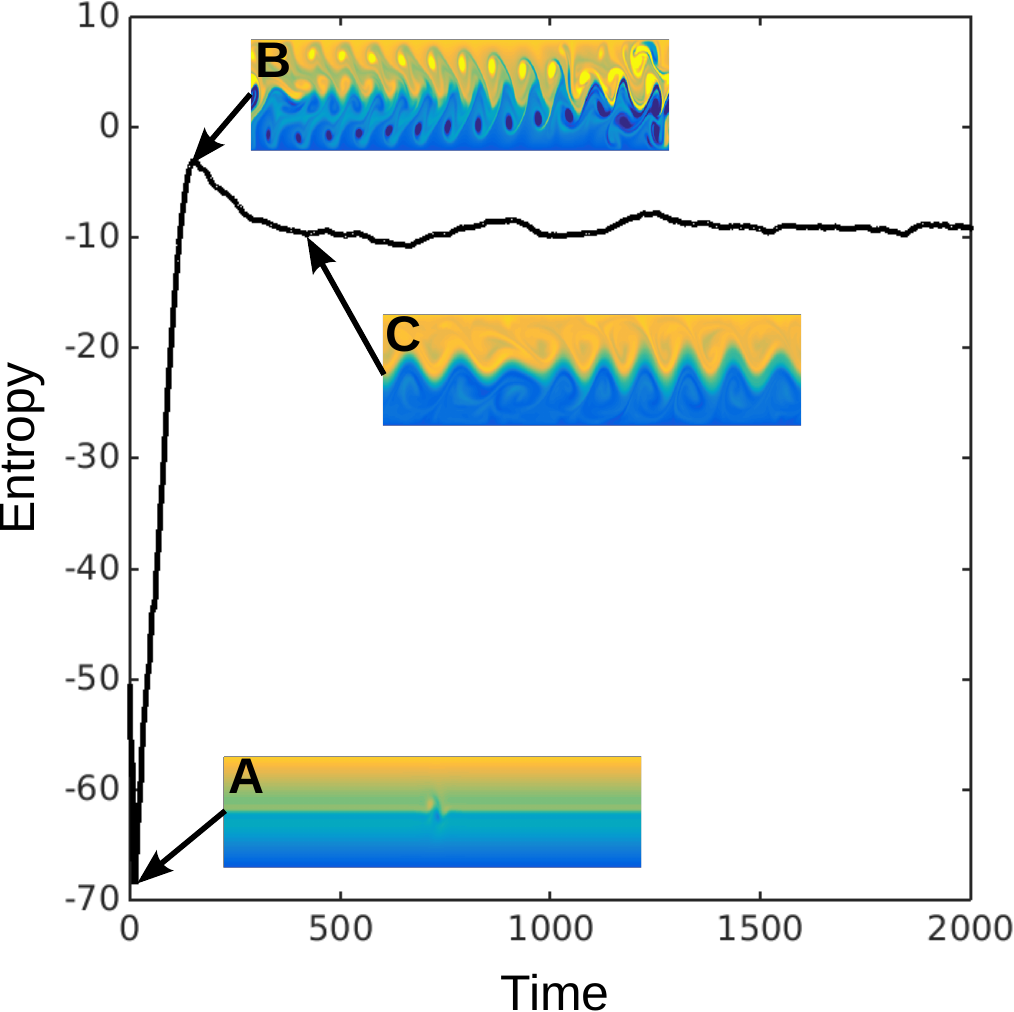}}
\caption{Evolution of entropy during spin-up of forced-dissipative flow. A - Initially laminar flow with small perturbation. As the flow becomes unstable the eddy-mixing entropy increases rapidly. B - Entropy reaches a global maximum at the point where turbulence has spread across the whole domain. Disorder is at small scales. C - Coincident with the emergence of a large scale Rossby wave the entropy decreases. Subsequently the entropy fluctuates around a balanced time-mean value.}
\label{exentsnap}
\end{figure}
\ref{exentsnap} shows the evolution of entropy as the system evolves to a statistically steady state. Initially, in region A, the flow is near laminar with only the small initial perturbation. We see that this corresponds to a low value of entropy. Once instabilities begin to grow the corresponding growth of entropy is fast and grows to a maximum value much like the evolution in the freely-decaying simulations. At the maximum of entropy, region B, the turbulence has covered the whole domain with small scale eddies. As these eddies mix the potential vorticity we see a slump in the entropy. When we examine the flow at the bottom of the slump, region C, we see that a large scale Rossby wave has emerged propagating on a sharp potential vorticity gradient corresponding to a mixing barrier. This decrease of entropy, or disorder, in the system allows us to describe the way in which energy has condensed at large scales in an entropic sense. The concomitance of this decrease in entropy with the emergence of large scales leads to a novel interpretation of well known inverse cascade phenomena: the emergence of large scales coherent can be described as the decrease of entropy in this system. 

At longer times the entropy fluctuates around a balanced steady state value. This behaviour of entropy is the same for all the forced-dissipative simulations, $\mathrm{FD_1,...,FD_7}$, shown in Figure
\begin{figure}
\centerline{\includegraphics[width=8cm]{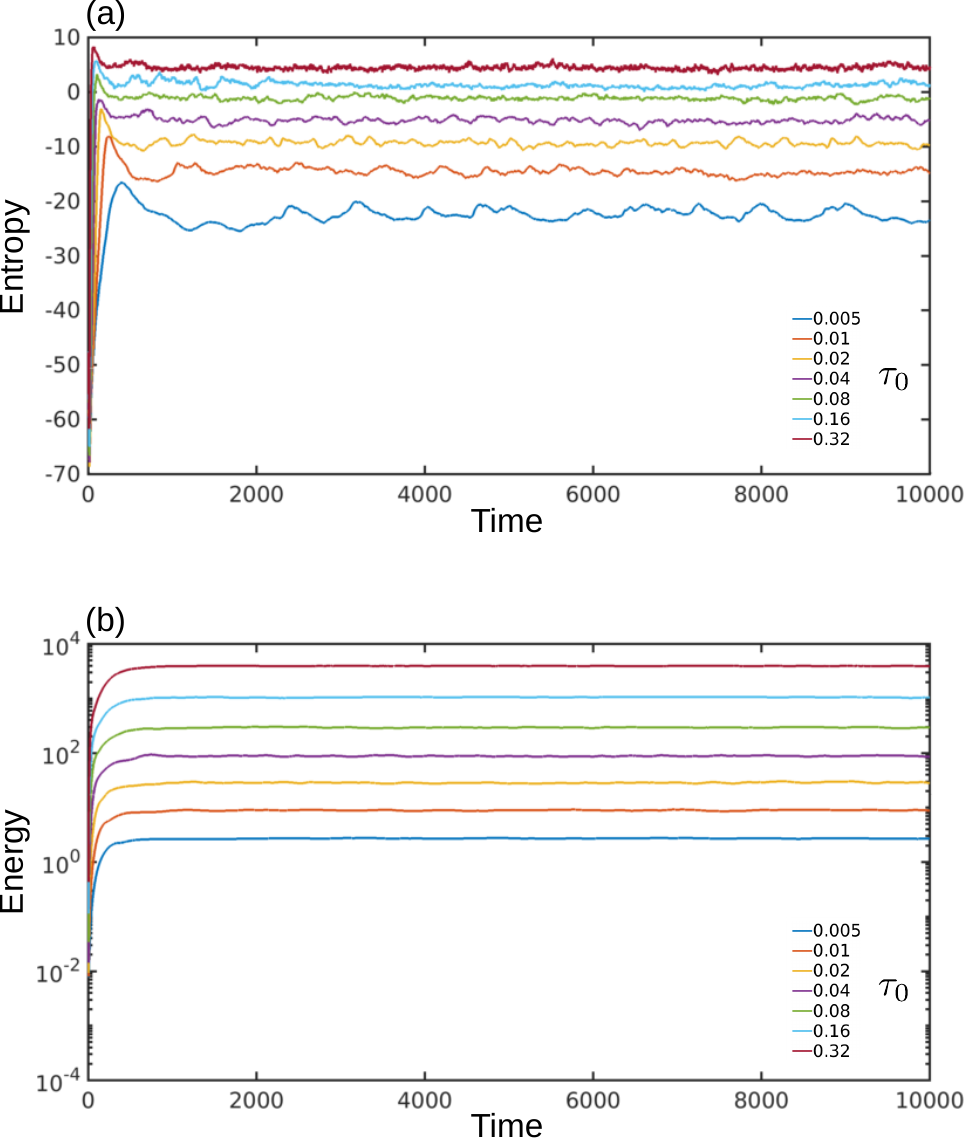}}
\caption{(a) Evolution of entropy in simulations $\mathrm{FD_1,...,FD_7}$. We see that entropy behaves as a balanced quantity in a statistically steady state: in the time mean the sources of entropy are equal to the sinks of entropy. (b) Evolution of energy in simulations $\mathrm{FD_1,...,FD_7}$ shown on a logarithmic scale. Like entropy the energy is balanced in a statistically steady state. The energy takes a longer time than the entropy to reach steady state balance.}
\label{FDentropyenergy}
\end{figure}
\ref{FDentropyenergy}a, much like statistically steady state balance of energy, shown in Figure \ref{FDentropyenergy}b. Both the time-mean entropy and energy increase with the wind stress strength in steady state. The balanced steady state behaviour of the entropy is explained, according to the reasoning of Section \ref{analent}, by the competition between the advective production of entropy and the constant sink due to linear drag, that is
\begin{equation}
\overline{P}-r = 0,
\end{equation} 
where the over over-line denotes the time-mean in statistically steady state. It is important to note that, because $-r$ is a merely a negative number, both the increase and decrease in the entropy fluctuations arise from the advective production, $P$. That is, \emph{eddies can act as both a source and a sink of entropy}. It is important to note that the action of $P$ as a source and a sink must be associated with the presence of dissipation which allows there to be fluctuations in otherwise conserved global quantities such as energy and entropy which exhibit inverse and direct transfers between scales. 

Although the time derivative of entropy has no explicit dependence on forcing, the forcing does supply energy to the turbulent motions by sustaining the eddy production of entropy, unlike in the case of the freely-decaying experiment. The forcing implicitly sets the maximum and steady state value of entropy, we will further consider how the entropy is related to other well-known dynamical quantities in the following section.  

\section{Relation to the maximum entropy principle}
\label{maxent}
\subsection{Time evolution of entropy and the maximum entropy principle}
In what has been discussed so far we have considered the derivative of entropy with respect to time. Now we consider its relation to the maximum entropy principle which is at the core of the equilibrium statistical mechanics theory of ideal geophysical flow \citep{jungetal2006}. 

The maximum entropy principle states that the entropy should be stationary with respect to variations in the probability distribution, $\rho$, given appropriate dynamical constraints. We can relate the time derivative of the entropy, $S$, to the functional derivative using the relation
\begin{equation}\label{dtimetodrho}
\diff{S}{t} = \int \ndint{2}{x} \mathrm{d}\tilde{q} \, \partd{\rho}{t} \funcd{S}{\rho}.
\end{equation}
Assuming that the system is in a maximum entropy state constrained by the value of energy and $N$ Casimirs we have that the variational problem
\begin{eqnarray}\label{variation}
\funcd{S}{\rho} &+& \alpha(t) \funcd{}{\rho}\left(\int \ndint{2}{x} \mathrm{d}\tilde{q} \, \langle \psi \rangle \tilde{q} \rho  - E(t) \right) \nonumber \\
&+& \sum_{n=1}^{N} \gamma_n(t) \funcd{}{\rho} \left(\int \ndint{2}{x} \mathrm{d}\tilde{q} \, \tilde{q}^n \rho - C_n(t)\right) = 0,
\end{eqnarray}
is satisfied, where $\alpha$ and $\gamma_n$s are Lagrange multipliers and where $E(t)$ and $C_n(t)$ are the time varying values of energy and Casimirs. Substituting this condition into \eqref{dtimetodrho}, we relate the time derivative of entropy to that of energy and the Casimirs:
\begin{equation}\label{sdoteandc}
\diff{S}{t} = - \alpha(t) \diff{E(t)}{t} - \sum_{n=1}^{N} \gamma_n(t) \diff{C_n(t)}{t}.
\end{equation}
This equation says that if the entropy is maximal under some constraints then the evolution of the entropy can be entirely explained by the evolution the quantities constraining it. We can split the time evolution of the Lagrange multipliers into temporal mean and fluctuations such that $\alpha = \overline{\alpha} + \alpha'$, and similarly for the other Lagrange multipliers. Assuming that the deviations in the Lagrange multipliers are small, which we will address in more detail in due course, and integrating \eqref{sdoteandc}, we can write an approximate relation for the time evolution of entropy in terms of the time evolution of energy and the Casimirs, giving
\begin{equation}\label{ssentevo}
S(t) \approx - \overline{\alpha} E(t) - \sum_{n=1}^{N} \overline{\gamma_n} C_n(t) + K,
\end{equation}
where $K$ is a constant of integration. This expression relies on two assumptions. Firstly, the entropy is maximized constrained by the value of energy and the Casimirs at any point in time; in other words, the system is in a quasi-equilibrium state where the time scale for changes in the balanced quantities is larger that the time the eddies take to drive the system to equilibrium. Secondly, the fluctuations in the Lagrange multipliers (sensitivities of the entropy) are small. We now turn to testing the relation, (\ref{ssentevo}), in order to test these assumptions.  

\subsection{Reconstruction of entropy evolution}
\label{recentevo}
We can test equation \eqref{ssentevo} by regressing the diagnosed time evolution of entropy onto the time evolution of energy and the other conserved quantities, the Casimirs, and comparing the reconstructed entropy time series against the diagnosed entropy time series. It was found the this procedure does not give a good reconstruction when the spin up is included in the time series, however there is good agreement when we only consider the statistically steady state: it is likely that the time dependence of the Lagrange multipliers is large during spin up but not in the statistically steady state. 

An example is given in Figure
\begin{figure*}
\centerline{\includegraphics[width=\textwidth]{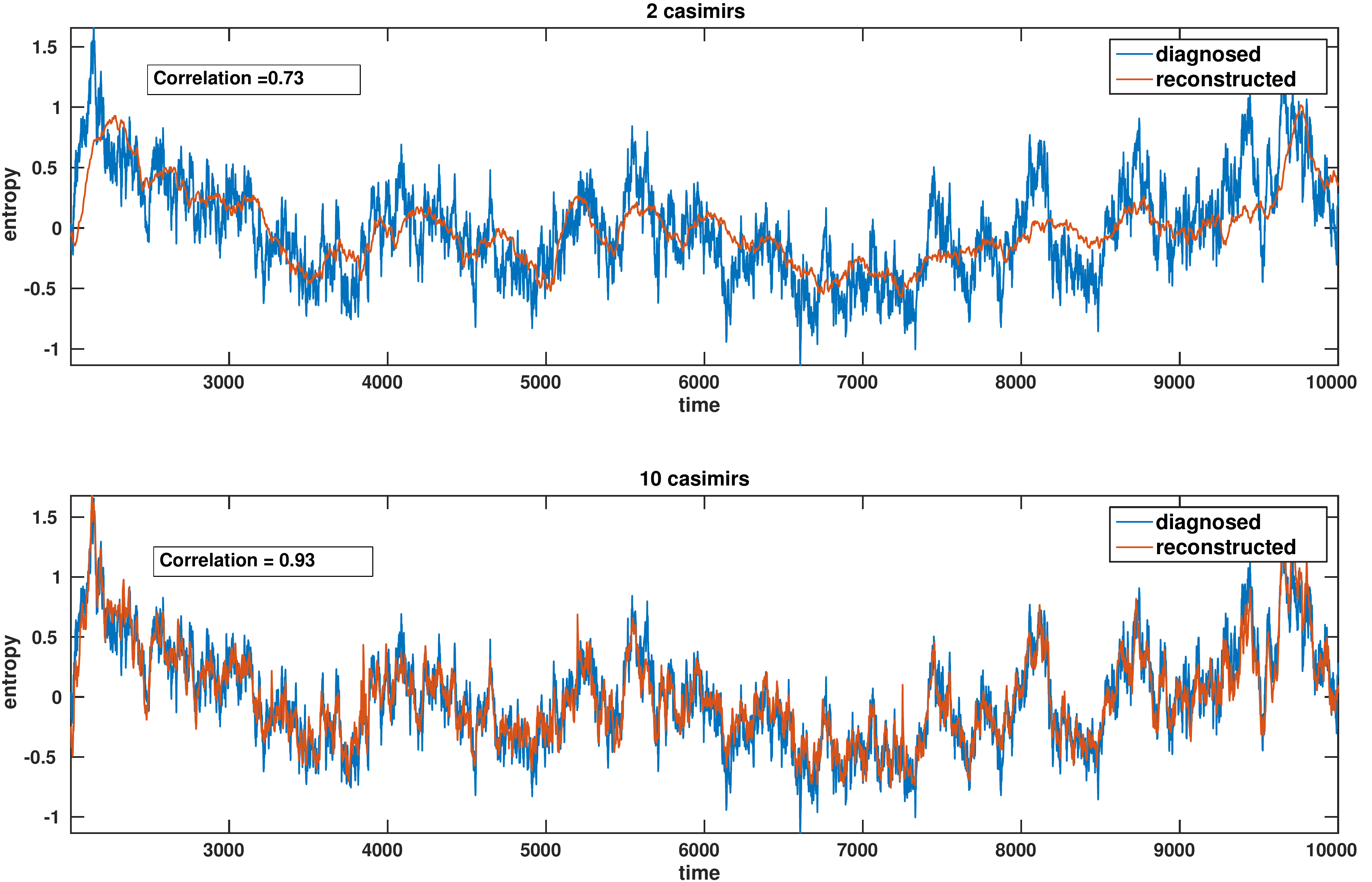}}
\caption{Reconstruction of the entropy evolution using relation \eqref{ssentevo}. \emph{Top} - using only the first two Casimirs, circulation and enstrophy. \emph{Bottom} - using first ten Casimirs. Both show significant correlation with the two Casimir reconstruction clearly matching the low frequency fluctuations, however, fails to capture magnitude of correlation. The ten Casimir reconstruction performs substantially better.} 
\label{entrec}
\end{figure*}
\ref{entrec} for simulation $\mathrm{D_6}$. Figure \ref{entrec} shows the reconstruction of the entropy time series using only the first two Casimirs, circulation and enstrophy, in addition to the energy as well as using Casimirs up to tenth order. The correlation for the first two Casimirs is 0.73, and when ten Casimirs are used the correlation is 0.93. This is a striking agreement and provides evidence that the turbulence acts to maximize entropy, according to equation \eqref{ssentevo}, at each point in time in statistically steady state. 

The analysis has been repeated for all forced-dissipative simulations $\mathrm{FD_1,...,FD_7}$ and including differing numbers of Casimirs. Figure 
\begin{figure*}
\centerline{\includegraphics[width=\textwidth]{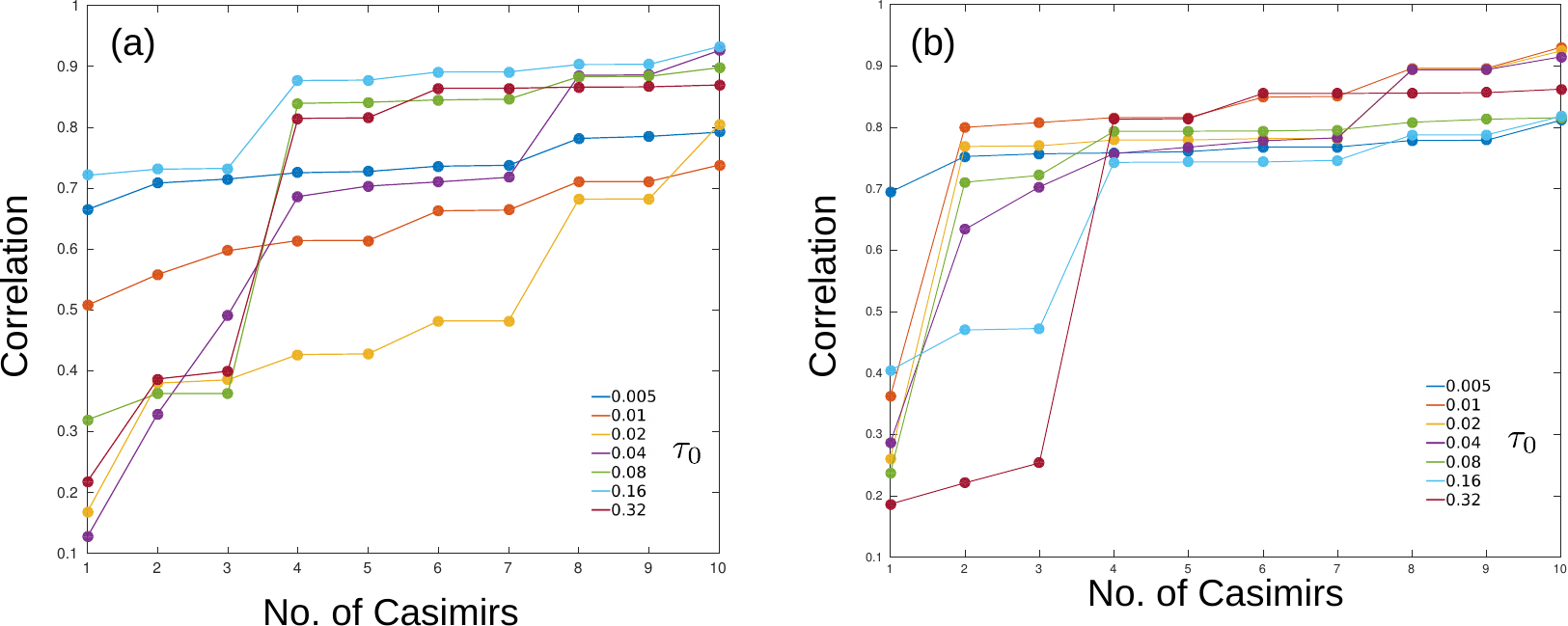}}
\caption{(a) Correlation between the diagnosed and reconstructed entropy time series following the relationship \eqref{ssentevo}. Entropy is calculated on zonal bands for simulations $\mathrm{FD_1,...,FD_7}$. We see a clear improvement with number of Casimirs included. This improvement is quantitatively different for each simulation. (b) Correlation between the diagnosed and reconstructed entropy time series following the relationship \eqref{ssentevo}. Entropy is calculated on instantaneous streamfunction for simulations $\mathrm{FD_1,...,FD_7}$. We see a clear improvement with number of Casimirs included. This improvement is quantitatively different for each simulation but now all simulations show high correlation of greater than 0.7 with only four Casimirs. This is to be constrasted with a).}
\label{correntreccombo}
\end{figure*}
\ref{correntreccombo}a shows the correlation between reconstructed and diagnosed entropy time series as a function of number of Casimirs for $\mathrm{FD_1,...,FD_7}$. We see a marked increase in the correlation with increased numbers of Casimirs. This shows the importance of higher order Casimirs in this statistical mechanics approach. It is also apparent that odd power Casimirs do not contribute significantly to increasing the correlation. 

We can repeat this analysis, now calculating the entropy along instantaneous streamfunction contours. This means that the macro-cells defined in Section \ref{entdef} become contours of streamfunction rather than zonal bands as has been used up to now. The entropy will change as a consequence of this transformation but the energy and Casimirs will not - this implies that only the projection in \eqref{ssentevo} of the entropy evolution onto the energy and Casimirs that will change. The correlation of the reconstructed entropy with the diagnosed entropy for this choice of macro-cell is shown in Figure \ref{correntreccombo}b. 

The main point to note here is that the correlation is higher for fewer Casimirs. The fourth order Casimir seems to be of particular importance with all simulations having a correlation of greater than 0.7 if four Casimirs are used. This agrees with the observation of \cite{davidetal2017a} that viewing potential vorticity distributions in a more of a  Lagrangian sense acts to simplify the statistics requiring a reduced number of moments to describe to probability distribution. The particular importance of the fourth order Casimir here is in contrast to the study of \cite{abramovmajda2003} who found that, in their numerical set-up, no Casimir above order three was relevant; this indicates that the particular Casimirs which are most important depend heavily on the particulars of any given system (i.e. domain geometry and nature of forcing). 

The fact that it is possible, to a large extent, to reconstruct the statistically steady state time series for entropy using the corresponding dynamically balanced quantities of the flow is tantalizingly suggestive that in a statistically steady state of a forced-dissipative flow the eddies push the system into a quasi-equilibrium. That is, although not in a true equilibrium (ideal flow with maximal entropy), the rate at which eddies push the entropy to its maximum allowed value is much faster than the time scales over which the conserved quantities are changing. 

As we saw in Sections \ref{freedecay} and \ref{force-diss}, the eddies can act as a source or sink of entropy. Indeed, the fact that the balanced quantities such as energy (Figure \ref{FDentropyenergy}b) fluctuate at all is due to the turbulence, if there were no non-linearities then we would have steady flow and no fluctuations. We suggest that the eddies play a double role, simultaneously maintaining the quasi-equilibrium and modulating its constraints. Further work, over a wider range of parameters, is required to obtain firmer evidence for the maximum entropy principle at work. 

\subsection{Solving for the Lagrange multipliers}

Solving the variational problem, (\ref{variation}), gives us a probability distribution in terms of a set of unknown Lagrange multipliers, in this section we describe the method for determining these from knowledge of the energy and Casimirs of the flow. To determine the Lagrange multipliers it is necessary to solve the non-linear simultaneous equations
\begin{equation}
- 2 E = - \partd{}{\alpha} \int \ndint{2}{x} \, \ln \mathcal{Z}(\alpha,\gamma_1,...,\gamma_N)
\end{equation}
for the energy constraint and 
\begin{equation}
C_n = - \partd{}{\gamma_n} \int \ndint{2}{x} \, \ln \mathcal{Z}(\alpha,\gamma_1,...,\gamma_N)
\end{equation}
for each Casimir constraint. Here $\mathcal{Z}$ is the local normalization, or the partition function, of the probability distribution given as
\begin{equation}
\mathcal{Z} = -\int \mathrm{d}\tilde{q} \, \exp \left(-\alpha \langle \psi \rangle \tilde{q} - \sum_{i=n}^{N} \gamma_n \tilde{q}^n \right).
\end{equation}
This problem is numerically difficult and its solution is not tackled in this study. However, we can proceed by reducing the dimensionality of the problem. Ironically this is achieved by first considering the case of infinite dimensions. Constraining the entropy of the flow the first $N$ polynomial Casimirs of the flow is a truncated version of the exact constraint. To constrain by \emph{all} Casimirs of the flow we constrain by the global potential vorticity distribution discussed in Section \ref{modelexp}. The constraint is given by 
\begin{equation}\label{Piintrho}
\Pi(\tilde{q}) = \int \ndint{2}{x} \, \rho(\tilde{q}|\mathbf{x}),
\end{equation}
and the Lagrange multiplier becomes a function of $\tilde{q}$, $\gamma(\tilde{q})$.  The corresponding variational problem now gives the solution 
\begin{equation}\label{maxentpdf}
\rho(\tilde{q}|\mathbf{x}) = \frac{1}{\mathcal{Z}} \exp \left( -\alpha \langle \psi \rangle \tilde{q} - \gamma(\tilde{q}) \right),
\end{equation}
for the probability distribution. Substituting \eqref{maxentpdf} into \eqref{Piintrho} we obtain the expression
\begin{equation}
\Pi(\tilde{q}) = e^{-\gamma(\tilde{q})} \int \ndint{2}{x} \, \frac{e^{-\alpha \langle \psi \rangle \tilde{q}}}{\mathcal{Z}}.
\end{equation}
The integral here is a function of potential vorticity only and we can write $\gamma(\tilde{q})$ in terms of $\Pi$ and the integral. This allows us to eliminate the Lagrange multiplier corresponding to the Casimir constraint leaving us with only $\alpha$ to find. Eliminating $\gamma$ from \eqref{maxentpdf}, we obtain the probability distribution in terms of the Lagrange multiplier, $\alpha$, and the global distribution, $\Pi$:
\begin{equation}\label{maxentpdf2}
\rho(\tilde{q}|\mathbf{x}) = \frac{1}{\mathcal{Z}(\mathbf{x}) \int \ndint{2}{x} \, \left(\frac{e^{-\alpha \langle \psi \rangle \tilde{q}}}{\mathcal{Z}(\mathbf{x})}\right)}\Pi(\tilde{q}) e^{-\alpha \langle \psi \rangle \tilde{q}}.
\end{equation}
A numerical method can be constructed for the two-dimensional problem of optimizing the values of $\alpha$ and $\mathcal{Z}$ simultaneously thus finding the maximum entropy distribution from only knowledge of global quantities. The results of this methodology follow.

\subsection{Reconstruction of statistics}

In this section we reconstruct the statistics from the maximum entropy distribution, (\ref{maxentpdf2}), by optimizing for the Lagrange multiplier, $\alpha$. The resulting values for $\alpha$ in simulations $\mathrm{FD_1,...,FD_7}$ are shown in Figure 
\begin{figure*}
\centerline{\includegraphics[width=\textwidth]{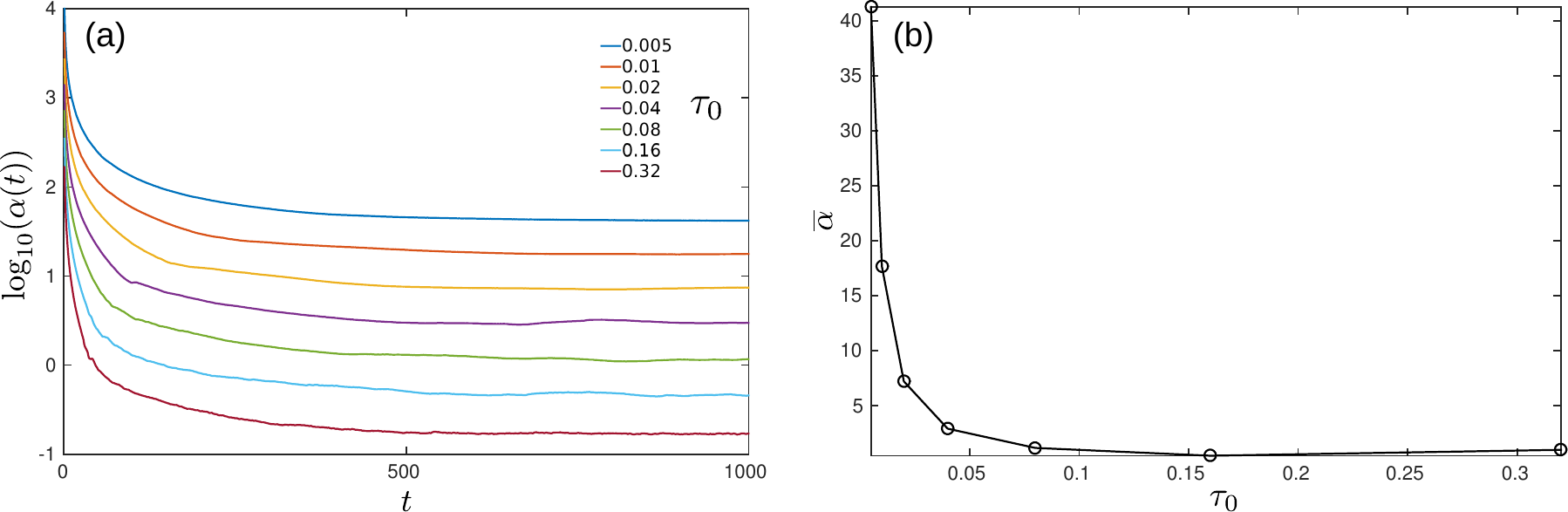}}
\caption{(a) Shows the evolution of $\log_{10} \alpha$ with time. We see a very quick relaxation to a steady state value with fluctuation around this value in steady state being small. (b) Shows the steady state mean value of alpha against the strength of the wind stress. We see at low forcing have a very high sensitivity of entropy to energy while at high forcing the sensitivity is low.}
\label{alphas}
\end{figure*}
\ref{alphas}. We see that the value of $\alpha$ has a strong dependence on the strength of wind stress and time. Figure \ref{alphas}a shows the evolution of $\alpha$ during spin up, we see a strong reduction in the value of $\alpha$ at short times with the value of $\alpha$ fluctuating about a statistically steady state value for long times. In steady state the fluctuations around the time-mean value is very small, this supports the assumption made in Section \ref{recentevo} to derive equation (\ref{ssentevo}). Also shown is the dependence of the time-mean value of $\alpha$ on the wind stress strength in Figure \ref{alphas}b. The steady state sensitivity of entropy to energy is drastically decreased with the strength of wind stress suggesting that the entropy of the system becomes insensitive to perturbations in energy at high wind stress.

Interpretations of the Lagrange multiplier, $\alpha$, come with a caveat; the numerically determined value of $\alpha$ is only as accurate as the maximum entropy hypothesis and, in particular, the mean field approximation for the energy. To test the accuracy of these devices we compare the reconstructed statistics from the distribution, (\ref{maxentpdf2}), and the diagnosed statistics from the numerical simulations. 

Figure
\begin{figure}
\centerline{\includegraphics[width=7cm]{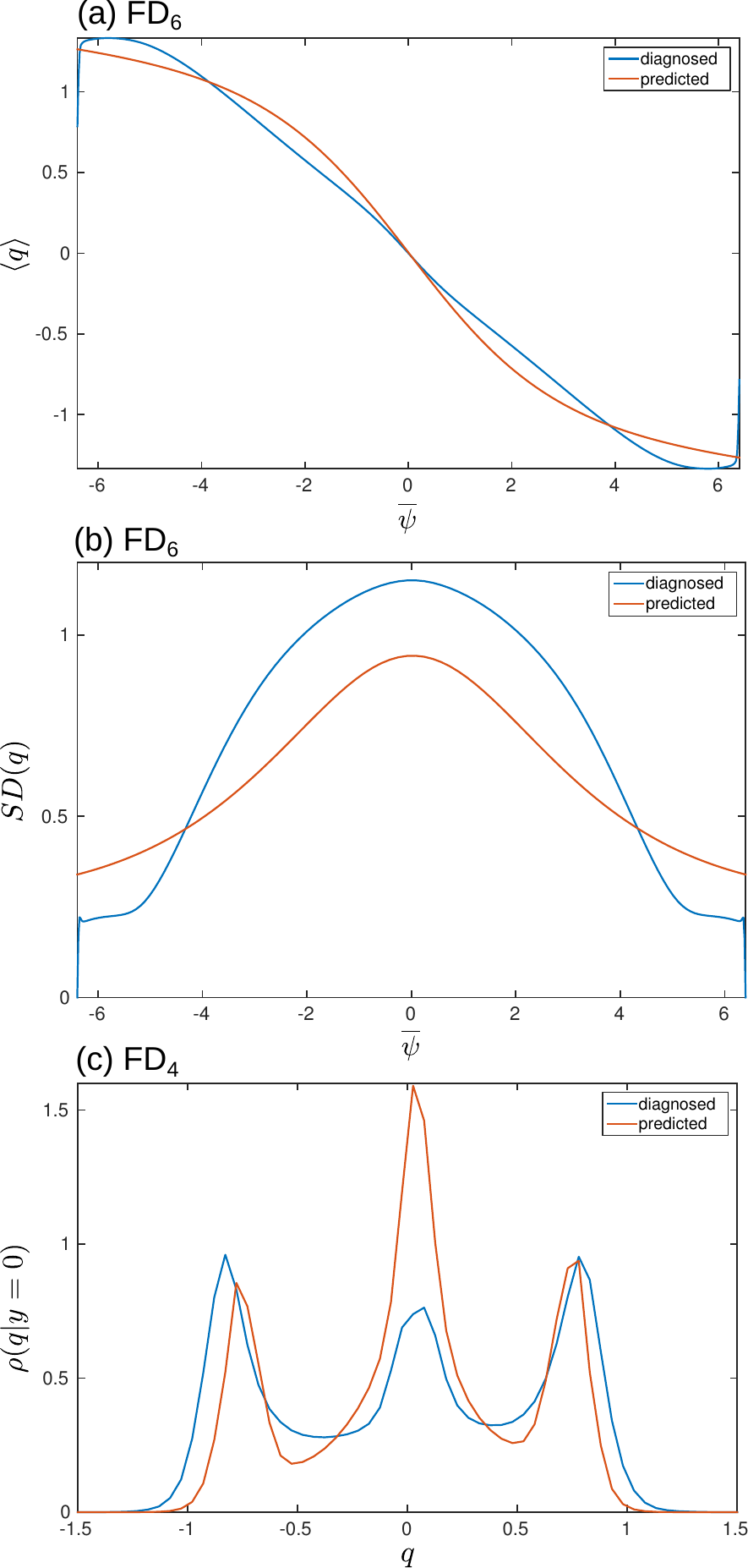}}
\caption{Diagnostics testing reconstructed maximum entropy statistics. (a) Comparison of the time-averaged mean potential vorticity against time-averaged streamfunction for simulation $\mathrm{FD_6}$. (b) Comparison of the time-averaged standard deviation of potential vorticity against time-averaged streamfunction for simulation $\mathrm{FD_6}$. (c) Comparision of the maximum entropy probability distribution and the diagnosed distribution from simulation $\mathrm{FD_6}$, distributions are evaluated for the centre of the channel, $y=0$.}
\label{recpdfdiag}
\end{figure}
\ref{recpdfdiag} shows diagnostics comparing the maximum entropy distribution evaluated using (\ref{maxentpdf2}) with the statistics diagnosed from the numerical simulation itself, that is, the `truth'. In short, despite displaying qualitative agreement with the simulations, the maximum entropy distribution fails to quantitatively capture the statistics of simulations $\mathrm{FD_1,...,FD_6}$. Figure \ref{recpdfdiag}a shows a comparison between the reconstructed and diagnosed $\overline{\langle q \rangle}-\overline{\langle \psi \rangle}$ relation for simulation $\mathrm{FD_6}$. The maximum entropy reconstruction shows good qualitative agreement but seems smoothed compared to the diagnosed relationship and fails to capture all the inflection points. Also for simulation $\mathrm{FD_6}$, Figure \ref{recpdfdiag}b compares the $\overline{\mathrm{SD}(q)}-\overline{\langle \psi \rangle}$ relation and we see that the quantitative agreement is poorer. The standard deviation is underestimated in the centre of channel while being overestimated in the flanks. 

Figure \ref{recpdfdiag}c compares the probability distribution, predicted and diagnosed, in the centre of the channel for simulation $\mathrm{FD_6}$. Despite capturing the trimodal nature of the distribution we can see that the maximum entropy distribution overestimates the weight of the central peak compared to the side peaks. As well as these comparisons, higher order moments, not shown here for brevity, were also considered such as the skewness and kurtosis. Again, despite giving qualitatively consistent features the skewness and kurtosis displayed considerable quantitative deviations from the simulations and were, for example, highly overestimated in the flanks. Many of the ways in which the maximum entropy statistics deviates from the simulations can be interpreted as a degradation of the strong persistent mixing barrier which the simulations $\mathrm{FD_1,...,FD_6}$ exhibit. The statistical nature and consequences of this mixing barrier has been extensively studied, for this model, in \cite{davidetal2017a}.  
 
In summary, the maximum entropy distribution, (\ref{maxentpdf2}), fails to quantitatively reproduce the simulated statistics despite the suggestive evidence for the entropy being maximized presented in Section \ref{recentevo}. The derivation of the maximum entropy distribution relies on two main assumptions in the equilibrium, Miller-Robert-Sommeria, statistical mechanics: a) maximization of entropy; and b) the mean field approximation. We argue that the failure of the maximum entropy distribution presented in this numerical experiment can be attributed to the break-down of the mean field approximation in a forced dissipative system. This will be discussed in detail in the following section.

\section{The mean-field approximation}
\label{discmeanfield}

A possible reason for the lack of success of the predicted probability distribution function, (\ref{maxentpdf2}), is that the maximum entropy principle is not at work. Nevertheless, we believe that the analysis presented in Section \ref{recentevo} provides sufficient evidence to look for other reasons why (\ref{maxentpdf2}) fails to quantitatively capture the statistics. In this section we consider the mean field approximation of Miller-Robert-Sommeria equilibrium statistical mechanics. The energy of the flow is given as
\begin{equation}
E[q] = -\frac{1}{2} \int \mathbf{\mathrm{d}x} \, \psi(\mathbf{x}) (q(\mathbf{x}) - \beta y), 
\end{equation}
because of the presence of $\psi$ it is not clear how make the necessary substitution to write $E$ as a functional of $\rho$ allowing us to tackle this constraint analytically. We rewrite the energy as
\begin{equation}
E[q]= -\frac{1}{2}\int \int \mathbf{\mathrm{d}x}  \mathbf{\mathrm{d}x'} \, (q(\mathbf{x}) - \beta y) G(\mathbf{x},\mathbf{x}') q(\mathbf{x}')
\end{equation}
where $G(\mathbf{x},\mathbf{x}')$ is the Green's function of the differential operator defined by, $q = \nabla^2 \psi +\beta y$. Now, by swapping the potential vorticity field for its average value, and defining $\langle \tilde{q} \rangle \equiv \nabla^2\langle \psi \rangle+\beta y$, we get;
\begin{equation}\label{enecon}
E_M[\rho] = -\frac{1}{2} \int \mathbf{\mathrm{d}x} \, \langle \psi \rangle (\langle \tilde{q} \rangle - \beta y).
\end{equation}
This step is the \emph{mean-field approximation}, often used in models of condensed matter physics. In essence, we are saying that, rather than considering the interaction energy between all pairs of potential vorticity patches, we consider that each patch of potential vorticity only feels mean effect of all other patches. 

For an ideal fluid in statistical equilibrium, the mean-field approximation ceases to be approximate and becomes exact due to the non-local aspect of the Green's function, $G$ \citep{bouchetvenaille2012}. However, this necessitates two other properties of ideal equilibria: a) that the mean eddy potential vorticity flux is zero, $\nabla \cdot \langle \mathbf{u}'q' \rangle = 0$; and b) that neighbouring macro-cells of the flow are uncorrelated. These properties are fundamentally linked with the mean field approximation and are manifestly not satisfied in a forced-dissipative statistically steady state. Therefore, we suggest that while the maximization of entropy might be an useful organizing principle in forced-dissipative flow, the mean-field approximation remains only a crude approximation. We propose two potential avenues for future study to tackle this problem.
\begin{itemize}
\item It may be possible to find a coarse-graining (i.e., macro-cells) which is partially Lagrangian to reduce difference between  $E[q(\mathbf{x})]$ and $E_M[\rho(\tilde{q}|\mathbf{x})]$. This is the approach taken in \cite{jungetal2006} which produces good agreement between experiment and equilibrium theory by moving into a frame of reference moving at the phase speed of a large scale Rossby wave. However, this is unlikely to work in the presence of multiple wave modes as is the case with simulations $\mathrm{FD_1,...,FD_6}$, see discussion in \cite{davidetal2017a}.
\item Alternatively, we speculate that a perturbation method could be applied to the mean-field approximation to yield a more realistic probability distribution. However, precisely how this might be achieved remains to be determined and is subject of future work.
\end{itemize}  
Indeed, it may be that a combination of the two approaches will provide the means of deriving forced-dissipative statistics from the maximum entropy principle.

\section{Conclusion}
\label{conc}

In this study we have shown how an eddy mixing entropy can be used as a measure of turbulent disorder. By deriving the influence of forcing and linear drag, we were able to use entropy to describe the turbulence in a freely-decaying and forced-dissipative flow. The evolution of entropy describes the three stages of the eddy life cycle and eddy-mean interaction: growth of instability, formation of large scale coherent structures and steady state fluctuations. In particular, the eddy production of entropy, which has been focus of much theoretic inquiry, can be explicitly computed from data. This will inform work on stochastic parametrization by describing the disorder in a turbulent jet in a way that links to both statistical physics and information theory.

The relationship between the temporal evolution of entropy and the maximum entropy principle was considered in Section \ref{maxent}. Under the assumption of maximum entropy it was found that the time evolution of entropy was set by the time evolution of its constraints. Suggestive evidence was found that the entropy is maximized in the model simulations considered in this study. It is clear that if a variational problem can be used to infer the statistics then the number of Casimir constraints has to be large. 

With this evidence for the maximum entropy principle being a physically meaningful candidate for describing the behaviour of turbulence in the system studied here, we considered the problem of inferring the sub-grid scale statistics. This is equivalent to inferring the Lagrange multipliers, used in the maximum entropy variational problem, from the constraints applied. We presented the mathematical formulation of this problem in Section \ref{maxent} and showed how the dimensionality could be reduced given knowledge of the global potential vorticity distribution. Further, we reconstructed the maximum entropy statistics from knowledge of the energy, global potential vorticity distribution, and zonal mean streamfunction as a functions of time. We find that although the maximum entropy statistics reproduces qualitatively representative features of the flow, quantitative agreement is lacking, especially for higher order statistical moments. In Section \ref{discmeanfield}, the mean-field approximation was discussed as a potential culprit for the quantitative disagreement and avenues for future investigation were proposed.

In this study we have presented eddy-mixing entropy as both a descriptive tool and a dynamically balanced quantity in a barotropic turbulent jet. We have also demonstrated the relationship between the statistical mechanics of forced-dissipative flow and well-known globally balanced quantities such as the energy and enstrophy of the flow. In doing so we were able to provide evidence for the maximum entropy principle in a forced-dissipative system, where the eddies act to maximize entropy under time varying constraints. The question of the usefulness of statistical mechanics theories, such as the Miller-Robert-Sommeria theory, in understanding the statistically steady states of ideal two-dimensional and geophysical turbulence has received much attention \citep[e.g.][]{wangvallis1994,jungetal2006,qimarston2014,dritscheletal2015}. By explicitly considering the evolution of eddy mixing entropy in a forced-dissipative model we are able to demonstrate the importance and utility of eddy mixing entropy in the study of forced-dissipative geophysical turbulence, opening the door to revisiting the application of statistical mechanics to ocean mesoscale eddy parameterizations \citep[cf.][]{robertsommeria1992,kazantsevetal1998,chavanis2014}.    

\begin{acknowledgements}
This work is funded by the UK Natural Environment Research Council.
\end{acknowledgements}

\appendix*
\section{Numerical computation of entropy}

It is important to note the difference between the discrete, or Shannon, entropy
\begin{equation}\label{discent}
S = - \sum_{i} \rho_i \ln \rho_i,
\end{equation}
and the continuous, or differential, entropy we use in this study
\begin{equation}\label{content}
S = - \int \mathrm{d}x\, \rho(x) \ln \rho(x).
\end{equation}
One of the clear differences between these two entropies is that the continuous entropy can become negative whereas the discrete entropy is never less than zero. On the other hand the continuous entropy can be negative and indeed tends to negative infinity for the asymptotic limit of a delta-function. This means that we need to be careful when numerically evaluating an estimator for the continuous entropy. Na\"{i}vely using the standard discrete approximation for the integral in \eqref{content} leads to calculating a quantity proportional to the discrete entropy, \eqref{discent}. To find an approximation for, \eqref{content}, we must evaluate the quantity 
\begin{equation}
S \approx - \Delta x \sum_{i} \rho_i \ln \rho_i + \ln \Delta x,
\end{equation}
where $\rho_i$ becomes a histogram approximation to the distribution $\rho(x)$. However this method of approximation was found to be biased and introduced a systematic error into the results presented in this study. 

Instead, we used a sample-spacing estimator for the distribution leading to an improved numerical approximation for the continuous entropy \citep{vasicek1976,beirlantetal1997}. The sample-spacing estimator relies of the idea that when the data is ordered, from smallest to largest value, then the reciprocal of the difference between two samples, separated by $m$, spaces is an estimator for the probability density. Using this, the following expression is found for the entropy
\begin{equation}
S \approx \frac{1}{N-m} \sum_{i=1}^{N-m} \ln \left(\frac{N+1}{m}(x^{(i+m)} - x^{(i)}) \right).
\end{equation}
Here, $N$ is the number of samples; $m$ is the spacing size; and $x^{(i)}$ represents the $i^{\mathrm{th}}$ ordered sample. This method was used to evaluate the entropy throughout this study and is found to be considerably better than more na\"{i}ve methods. The simpler methods proved to have a strong dependence on the choice of histogram bin width rendering them unusable for quantitative comparison with theory. 
 
\bibliographystyle{apa}
\bibliography{./tdbib}

\end{document}